\documentclass[aps,preprint,showpacs,showkeys]{revtex4}
\usepackage{amssymb}

\input{tcilatex}

\begin{document}

\title{Integro-differential diffusion equation for continuous time random
walk }
\author{Kwok Sau Fa$^{1}$ and K. G. Wang$^{2}$}
\affiliation{$^{1}$Departamento de F\'{\i}sica, Universidade Estadual de Maring\'{a}, Av.
Colombo 5790, 87020-900, \ Maring\'{a}-PR, Brazil,}
\email{kwok@dfi.uem.br}
\affiliation{$^{2}$Department of Physics \& Space Sciences, Materials Science and
Nanotechnology Institute, Florida Institute of Technology, Melbourne, FL
32901, USA}

\begin{abstract}
In this paper we present an integro-differential diffusion equation for
continuous time random walk that is valid for a generic waiting time
probability density function. Using this equation we also study diffusion
behaviors for a couple of specific waiting time probability density
functions such as exponential, and a combination of power law and
generalized Mittag-Leffler function. We show that for the case of the
exponential waiting time probability density function a normal diffusion is
generated and the probability density function is Gaussian distribution. In
the case of the combination of a power-law and generalized Mittag-Leffler
waiting probability density function we obtain the subdiffusive behavior for
all the time regions from small to large times, and probability density
function is non-Gaussian distribution.
\end{abstract}

\maketitle

\section{\protect\bigskip Introduction}

Diffusion is a ubiquitous phenomenon, and it is one of the fundamental
mechanisms for transport of materials in physical, chemical and biological
systems. The well-known example of a diffusion process is the Brownian
motion. Diffusion processes are classified according to their mean-square
displacements. The normal diffusion is that the mean-squared displacement
grows linearly with time, and in other situations the processes are said to
exhibit anomalous diffusion. Nowadays, there are several approaches to
describe anomalous diffusion processes, and they can be applied to many
situations of natural systems \cite%
{kubo,risken,georges,klafter,klafter2,zasla,lee,westrev,Wang99}. One of the
most interesting features incorporated into these approaches is the memory
effect. In particular, the memory effect incorporated into the Langevin
approach, referred to as generalized Langevin equation (GLE) \cite{Wang99},
can be associated to the retardation of friction and fractal media \cite%
{west,west1}. Moreover, according to the fluctuation-dissipation theorem 
\cite{kubo}, the internal friction is directly related to the correlation
function of the random force.

In many situations, a finite correlated noise is necessary for describing
the real systems in equilibrium states. However, in order to describe
anomalous diffusion, a nonlocal friction should be employed so that it
satisfies the fluctuation-dissipation theorem. For instance, anomalous
diffusion processes have been observed in a variety of systems such as
bacterial cytoplasm motion \cite{cox}, conformational fluctuations within a
single protein molecule \cite{kou} and fluorescence intermittency in single
enzymes \cite{chaudhury}. These processes have been described by the GLE,
and the memory effect has also been shown in the generalized Fokker-Planck
equations (GFPE) \cite{Wang99}.

The continuous time random walk (CTRW) of Montroll and Weiss \cite%
{Montroll65} has been employed to describe anomalous diffusion \cite%
{Scher73, Shlesinger74, Scher75, klafter3, Kotulski95, Barkai97,coffey}. The
CTRW with a power-law waiting time probability density function (pdf) \cite%
{Balakrishnan85, Barkai00} was also linked to the following fractional
Fokker-Planck equation,

\begin{equation}
\frac{\partial \rho (x,t)}{\partial t}=_{0}D_{t}^{1-\alpha }K_{\alpha }\frac{%
\partial ^{2}}{\partial x^{2}}\rho (x,t)\text{ , }  \label{eq1}
\end{equation}%
where 
\begin{equation}
_{0}D_{t}^{1-\alpha }\rho (x,t)\text{ }=\frac{1}{\Gamma (\alpha )}\frac{%
\partial }{\partial t}\int_{0}^{t}\frac{\rho (x,t_{1})}{\left(
t-t_{1}\right) ^{1-\alpha }}\text{d}t_{1}\text{ }  \label{eq2}
\end{equation}%
is the Riemann-Liouville fractional derivative and $\Gamma (z)$ is the Gamma
function. $\rho (x,t)$d$x$ is the probability for finding particle in
position between $x$ and $x+dx$ at time t.

The CTRW model may be described by a set of Langevin equations \cite%
{klafter, foged1,foged2} or an appropriate generalized master equation \cite%
{klafter3,klafter4,berko}. The pdf $\rho (x,t)$ obeys the following equation
in Fourier-Laplace space \cite{klafter}:

\begin{equation}
\rho (k,s)=\frac{1-g(s)}{s}\frac{\rho _{0}(k)}{1-\psi (k,s)}\text{ ,}
\label{eq3}
\end{equation}%
where $\rho _{0}(k)$ is the Fourier transform of the initial condition $\rho
_{0}(x)$, $\psi (x,t)$ is the jump pdf and \ $g(t)$ is the waiting time pdf
defined by%
\begin{equation}
g(t)=\int_{-\infty }^{\infty }\psi (x,t)\text{d}x\text{ .}  \label{eq3a}
\end{equation}%
Moreover, from the jump pdf we also have the jump length pdf defined by%
\begin{equation}
\phi (x)=\int_{0}^{\infty }\psi (x,t)\text{d}t\text{ .}  \label{eq3b}
\end{equation}

The continuous time random walk model can be simplified through the
decoupled jump pdf $\psi (k,s)=\phi (k)g(s)$. Furthermore, jump length pdf
can be approximated as following \cite{klafter}:%
\begin{equation}
\phi (k)\sim 1-Dk^{2}+O(k^{4})\text{ .}  \label{eq4}
\end{equation}%
Different types of the CTRW models are specified through specifying the
waiting time pdf. The CTRW model is also connected to a class of
Fokker-Planck equations \cite{klafter,baule}. Furthermore, in the CTRW
model, solutions for $\rho (x,t)$ under the condition of long-tailed waiting
time pdf can be found in Refs. \cite{wang,weiss1,weiss2}.

Substituting Eq.(\ref{eq4}) into Eq.(\ref{eq3}), we have 
\begin{equation}
\rho (k,s)=\frac{1-g(s)}{s}\frac{\rho _{0}(k)}{1-\left( 1-Dk^{2}\right) g(s)}%
\text{ .}  \label{eq4a}
\end{equation}%
When Cakir et al. \cite{cakir} studied trajectory and density memory, the
Eq.(43) in their paper is same as our Eq.(\ref{eq4a}) with consideration of $%
\rho _{0}(k)=1$. In their study, the initial condition is that when $t=0$,
particle is at origin, and therefore $\rho _{0}(k)=1$. However, Eq. (\ref%
{eq4a}) can be applied to cases without requirement of $\rho _{0}(k)=1$ that
appeared in Cakir et al.' paper \cite{cakir}. In particular, the CTRW model
can be classified by the characteristic waiting time $T$ and the jump length
variance $\Sigma ^{2}$ defined by%
\begin{equation}
T=\int_{0}^{\infty }tg(t)\text{d}t\text{ ,}  \label{eq4b}
\end{equation}%
and%
\begin{equation}
\Sigma ^{2}=\int_{-\infty }^{\infty }x^{2}\phi (x)\text{d}x\text{ .}
\label{eq4c}
\end{equation}%
For finite $T$ and $\Sigma ^{2}$, the long-time limit corresponds to the
Brownian motion \cite{klafter}. Although Eq. (\ref{eq4a}) is valid for a
finite jump length variance, anomalous diffusion can be produced by Eq.(\ref%
{eq4a}) with appropriate choices of waiting time pdf.

The aim of this work is: (1) from Eq.(\ref{eq4a}) to investigate general
behavior of particle's diffusion without any specified waiting time pdf; (2)
study the effects of different waiting time pdf's on the behavior of
particle's diffusion in the framework of CTRW. This paper is organized as
follows. In Section 2 we study CTRW under a general waiting time pdf and two
specific waiting time pdf's. In Section 3, we show the exact solutions for
pdf $\rho (x,t)$. Finally, conclusions are presented in Section 4.

\section{Continuous time random walk and integro-differential equation for
general waiting time probability density function}

\bigskip

Eq.(\ref{eq4a}) was already employed to study diffusion behavior when the
form of waiting time pdf $g(t)$ was first specified such as exponential
function or a power-law function in the long-time limit \cite{klafter}.
However, for a generic form of $g(t)$ Eq.(\ref{eq4a}) is not convenient to
be used to study diffusion behavior due to the difficulty in the inverse
transformations of Fourier and Laplace in Eq.(\ref{eq4a}). In this paper, we
will derive an integro-differential equation for CTRW, which is valid for
general waiting time probability density function. The details of derivation
can be found in Appendix A. This integro-differential equation for CTRW is
written by 
\begin{equation}
\frac{\partial \rho (x,t)}{\partial t}-\int_{0}^{t}g\left( t-t_{1}\right) 
\frac{\partial \rho (x,t_{1})}{\partial t_{1}}\text{d}t_{1}=D\frac{\partial 
}{\partial t}\int_{0}^{t}g\left( t-t_{1}\right) \frac{\partial ^{2}\rho
(x,t_{1})}{\partial x^{2}}\text{d}t_{1}\text{ .}  \label{eq10}
\end{equation}

For the case of $g(t)=t^{\alpha -1}/\Gamma (\alpha )$, it is noted that the
Caputo fractional derivative appears on the left side of Eq. (\ref{eq10})
and the Riemann-Liouville fractional derivative appears on the right side of
Eq. (\ref{eq10}). The Caputo fractional derivative requires the
integrability of the derivative and contains the initial value of the
function. Therefore the Caputo fractional derivative is more restrictive
than the Riemann-Liouville fractional derivative is. It should be pointed
out that the use of these operators may lead to different behaviors
including unphysical behavior in different systems \cite{kwok,ryabov}. Eq. (%
\ref{eq10}) is derived from a well-defined physical process, and no
ambiguities and unphysical behaviors are expected. The left side of Eq. (\ref%
{eq10}) shows the variation of $\rho (x,t)$ with respect to time, which does
not depend only on the ordinary derivative operator but also the difference
between ordinary and non-local integral operators.

Our Eq. (10) is different from the fractional Fokker-Planck Eq. (\ref{eq1})
in the structure because the fractional Fokker-Planck Eq. (\ref{eq1}) does
not have the second term of the left side of Eq. (\ref{eq10}). Substituting $%
g(t)=t^{\alpha -1}/\Gamma (\alpha )$ into Eq. (\ref{eq10}) cannot directly
obtain Eq. (\ref{eq1}). In order to exclude the second term of the left side
of Eq. (\ref{eq10}), the asymptotic behavior for $g(t)$ must be carefully
treated.

We note that%
\begin{equation}
\frac{\text{d}\left\langle x^{2}\right\rangle }{\text{d}t}=\int_{-\infty
}^{\infty }x^{2}\frac{\partial \rho (x,t)}{\partial t}\text{d}x\text{ .}
\label{eq10a}
\end{equation}%
Substitute Eq. (\ref{eq10}) into (\ref{eq10a}) yields%
\begin{equation}
\frac{\text{d}\left\langle x^{2}\right\rangle }{\text{d}t}%
=\int_{0}^{t}g\left( t-t_{1}\right) \frac{\text{d}\left\langle
x^{2}\right\rangle }{\text{d}t_{1}}\text{d}t_{1}+D\frac{\partial }{\partial t%
}\int_{0}^{t}g\left( t-t_{1}\right) \int_{-\infty }^{\infty }x^{2}\frac{%
\partial ^{2}\rho (x,t)}{\partial x^{2}}\text{d}x\text{d}t_{1}\text{ .}
\label{eq10aa}
\end{equation}%
After integrating the second term of the right side of Eq.(\ref{eq10aa}) by
parts twice, we have 
\begin{equation}
\frac{\text{d}\left\langle x^{2}\right\rangle }{\text{d}t}%
=\int_{0}^{t}g\left( t-t_{1}\right) \frac{\text{d}\left\langle
x^{2}\right\rangle }{\text{d}t_{1}}\text{d}t_{1}+2D\frac{\partial }{\partial
t}\int_{0}^{t}g\left( t-t_{1}\right) \int_{-\infty }^{\infty }\rho (x,t)%
\text{d}x\text{d}t_{1}\text{ ,}  \label{eq10ab}
\end{equation}%
where we also consider that $\lim_{x\rightarrow \pm \infty }\rho (x,t)$ and
it decreases faster than $1/x$. The normalization of the pdf $\rho (x,t)$
requires that $\int_{-\infty }^{\infty }\rho (x,t)$d$x=1$. Finally, Eq.(\ref%
{eq10ab}) can be rewritten as 
\begin{equation}
\frac{\text{d}\left\langle x^{2}\right\rangle }{\text{d}t}%
=\int_{0}^{t}g\left( t-t_{1}\right) \frac{\text{d}\left\langle
x^{2}\right\rangle }{\text{d}t_{1}}\text{d}t_{1}+2D\frac{\partial }{\partial
t}\int_{0}^{t}g\left( t-t_{1}\right) \text{d}t_{1}\text{ .}  \label{eq10b}
\end{equation}%
Employing the Laplace transform in Eq. (\ref{eq10b}) we obtain%
\begin{equation}
\left\langle x^{2}\right\rangle _{L}=\frac{\left\langle x^{2}\right\rangle
_{0}}{s}+\frac{2Dg(s)}{s\left[ 1-g(s)\right] }\text{ }  \label{eq10c}
\end{equation}%
in Laplace space. We have shown that from Eq. (\ref{eq10}) even without the
previous knowledge of the waiting time pdf, the second moment of
displacement i.e. the diffusion behavior can also be obtained in Eq. (\ref%
{eq10c}). This is one of advantages of Eq.(\ref{eq10}). Therefore, Eq.(\ref%
{eq10c}) is very general and valid for any waiting time pdf.

In fact, different diffusion behaviors can be found when substituting
different waiting time pdf's into Eq.(\ref{eq10c}). Considering that the
waiting time pdf must be positive and normalized, there exist only a few
simple functions that can be used as waiting time pdf's. Now we study
diffusion behaviors in two specific cases of waiting time pdf's.

\textit{First case}: $g_{1}(t)=\lambda e^{-\lambda t}$.

In this case the characteristic waiting time is finite. The Laplace
transform of $g_{1}(t)$ is given by%
\begin{equation}
g_{1}(s)=\frac{\lambda }{\lambda +s}\text{ .}  \label{eq11}
\end{equation}%
Substituting Eq. (\ref{eq11}) into Eq. (\ref{eq10c}) we obtain%
\begin{equation}
\left\langle x^{2}\right\rangle =\left\langle x^{2}\right\rangle
_{0}+2D\lambda t\text{ .}  \label{eq11a}
\end{equation}%
Eq.(\ref{eq11a}) shows the normal diffusion behavior. It has also shown that
the generalized diffusion equation (\ref{eq10}) can reduce to describe the
normal diffusion process when the waiting time pdf is a simple exponential
function.

\textit{Second case: \ }$g_{2}(t)=\lambda _{\alpha }t^{\alpha -1}E_{\alpha
,\alpha }(-\lambda _{\alpha }t^{\alpha })$.

$E_{\mu ,\nu }(y)$ is the generalized Mittag-Leffler function defined by 
\cite{mainardi,erdely} 
\begin{equation}
E_{\mu ,\nu }(y)=\sum_{n=0}^{\infty }\frac{y^{n}}{\Gamma \left( \nu +\mu
n\right) }\text{ , \ }\mu >0\text{, \ }\nu >0\text{ .}  \label{eq13aa}
\end{equation}%
We restrict the parameter $\alpha $ to $0<\alpha \leq 1$ because $E_{\alpha
,\alpha }(-\lambda _{\alpha }t^{\alpha })$, in this interval, is completely
monotone for $t>0$ \cite{mainardi}. We note that $g_{2}(t)$ can also be
written as 
\[
g_{2}(t)=-\frac{\text{d}}{\text{d}t}E_{\alpha ,1}(-\lambda _{\alpha
}t^{\alpha })\text{.} 
\]%
This waiting time pdf has been used to link the CTRW model and fractional
master equation \cite{hilfer}. It has also been adopted as subordination
function \cite{ascolani,gorenflo}. Fulger et al. \cite{fulger} incorporated
this waiting time pdf into Monte-carlo simulation of in uncoupled
continuous-time random walks with a L\'{e}vy $\alpha $-stable distribution
of jumps. The L\'{e}vy $\alpha $-stable distribution of jumps used in \cite%
{fulger} has a diverging jump length variance in contrast to the finite jump
length variance given by Eq. (\ref{eq4}) in this paper.

The waiting time pdf $g_{2}(t)$ has the following asymptotic behavior%
\begin{equation}
g_{2}(t)\sim -\frac{1}{\lambda _{\alpha }t^{1+\alpha }\Gamma \left( -\alpha
\right) }\text{ ,}  \label{eq13}
\end{equation}%
which has a power-law behavior. The characteristic waiting time is
divergent. The Laplace transform of $g_{2}(t)$ is given by%
\begin{equation}
g_{2}(s)=\frac{\lambda _{\alpha }}{\lambda _{\alpha }+s^{\alpha }}\text{ .}
\label{eq13a}
\end{equation}%
Substituting Eq. (\ref{eq13a}) into Eq. (\ref{eq10c}) we obtain%
\begin{equation}
\left\langle x^{2}\right\rangle =\left\langle x^{2}\right\rangle _{0}+\frac{%
2D\lambda _{\alpha }}{\Gamma \left( 1+\alpha \right) }t^{\alpha }\text{ .}
\label{eq13b}
\end{equation}%
This result is the same as that one obtained in \cite{klafter} for long-time
limit using approximation of a long-tailed waiting time pdf. However, it
should be noted that Eq. (\ref{eq13b}) describes the subdiffusive processes
for the all time regions rather than only for long-time or short time
region. This point is different from the case of long-tailed waiting time
pdf in Ref. \cite{klafter}.

\section{Exact solution for probability density}

Generally speaking, using Eq.(\ref{eq3}) we can calculate the pdf in the
framework of CTRW model. However, it is hard to obtain the exact pdf from
Eq.(\ref{eq3}). Now, we can obtain the exact solutions for $\rho (x,t)$ for
the two waiting time pdf's in Section 2 from Eq.(\ref{eq4a}). In the
following, we consider the initial condition as $\rho _{0}(k)=1$. After we
do Fourier inverse for $\rho (k,s)$ in Eq.(\ref{eq4a}), we have%
\begin{equation}
\rho (x,s)=\frac{1-g(s)}{2\pi Dsg(s)}\int_{-\infty }^{\infty }\frac{e^{ikx}}{%
k^{2}+\frac{1-g(s)}{Dg(s)}}\text{d}k\text{ .}  \label{eq14}
\end{equation}%
We note that the denominator can have three different poles: $1-g(s)=0$
gives trivial result, $(1-g(s))/g(s)<0$ gives the poles on the real axis and 
$(1-g(s))/g(s)>0$ gives the poles on the imaginary axis. For the waiting
time pdf's given in the previous section we can restrict to the case of the
poles on the imaginary axis. Thus, the solution for $\rho (x,s)$ is given by%
\begin{equation}
\rho (x,s)=\frac{1}{2\pi \sqrt{D}s}\sqrt{\frac{1-g(s)}{g(s)}}e^{-\frac{%
\left\vert x\right\vert }{\sqrt{D}}\sqrt{\frac{1-g(s)}{g(s)}}}\text{ .}
\label{eq14a}
\end{equation}%
Before obtaining the exact solutions to $\rho (x,t)$ we can calculate the
second moment in Laplace space which yields%
\begin{equation}
\left\langle x^{2}\right\rangle _{L}=\frac{2Dg(s)}{s\left[ 1-g(s)\right] }%
\text{ .}  \label{eq14b}
\end{equation}%
This result is similar to (\ref{eq10c}), except the term of initial value $%
\left\langle x^{2}\right\rangle _{0}$. This indirectly shows that Eq.(\ref%
{eq14a}) is correct.

For the case of $g_{1}(t)=\lambda e^{-\lambda t}$ the solution for $\rho
(x,t)$ is given by%
\begin{equation}
\rho _{1}(x,t)=\frac{1}{\sqrt{4\pi D\lambda t}}e^{-\frac{x^{2}}{4\lambda Dt}}%
\text{ .}  \label{eq14c}
\end{equation}%
The pdf shows the Gaussian shape and it is exactly the solution of the
normal diffusion equation \cite{klafter,risken}.

For the case of $g_{2}(t)=\lambda _{\alpha }t^{\alpha -1}E_{\alpha ,\alpha
}(-\lambda _{\alpha }t^{\alpha })$ the solution for $\rho (x,t)$ is given by

\begin{equation}
\rho _{2}(x,t)=\frac{1}{\sqrt{4D\lambda _{\alpha }t^{\alpha }}}%
\sum_{n=0}^{\infty }\frac{\left( -1\right) ^{n}}{n!\Gamma \left( 1-\alpha 
\frac{1+n}{2}\right) }\left( \frac{x^{2}}{D\lambda _{\alpha }t^{\alpha }}%
\right) ^{\frac{n}{2}}\text{ .}  \label{eq16a}
\end{equation}%
This result is exactly the solution of the fractional diffusion equation (%
\ref{eq1}). The pdf (\ref{eq16a}) presents pronounced cusps in x-coordinate
for different times. This form is different from the Gaussian shape (\ref%
{eq14c}) which presents smooth shape in the same coordinate \cite{klafter}.

Eq. (\ref{eq16a}) can also be expressed in terms of the Wright function $%
\Phi _{\eta ,\delta }\left( y\right) $ \cite{mainardi2} in the following
form:%
\begin{equation}
\rho _{2}(x,t)=\frac{1}{\sqrt{4D\lambda _{\alpha }t^{\alpha }}}\Phi _{-\frac{%
\alpha }{2},1-\frac{\alpha }{2}}\left( -\frac{\left\vert x\right\vert }{%
\sqrt{D\lambda _{\alpha }}t^{\frac{\alpha }{2}}}\right) \text{ ,}
\label{eq16b}
\end{equation}%
where%
\begin{equation}
\Phi _{\beta ,\delta }\left( y\right) =\sum_{n=0}^{\infty }\frac{y^{n}}{%
n!\Gamma \left( \beta n+\delta \right) },\text{ \ \ }\beta >-1,\text{ }%
\delta \in \mathbf{C}\text{ .}  \label{eq16c}
\end{equation}

\section{Conclusion}

In this work we have investigated the CTRW model with the decoupled jump
pdf. We have derived the integro-differential equation (\ref{eq10}). From
Eq.(\ref{eq10}) we have derived diffusion behavior that is explicitly
related to the waiting time pdf. Eq.(\ref{eq10}) is more flexible to be used
than Eq.(\ref{eq4a}) because using Eq.(10) we can avoid Laplace and Fourier
transformations and their inverse operations, for instance, Eq. (\ref{eq10b}%
) is ready to be used for calculating the second moment numerically. Both
Eqs.(\ref{eq10}) and (\ref{eq10c}) are very general and valid for any
waiting time pdf. To our knowledge, Eqs.(\ref{eq10}) and (\ref{eq10c}) have
not been published elsewhere. In fact, Eq. (\ref{eq4a}) or Eq. (\ref{eq10})
describes random walk model without external force. However, interesting
physical problems involve external force. Generalization of Eq. (\ref{eq10})
to include the effect of external force is a natural step. In addition,
possible generalization of Eq. (\ref{eq10}), e.g., using the master equation
approach \cite{klafter}, to include the effect of external force would be
more convenient and appropriate than Eq. (\ref{eq4a}).

Using Eq. (\ref{eq10}) we have studied diffusion behaviors for two cases:
(1) in the case of $g_{1}(t)=\lambda e^{-\lambda t}$, a normal diffusion or
normal Brownian motion is found; (2) in the case of $g_{2}(t)=\lambda
_{\alpha }t^{\alpha -1}E_{\alpha ,\alpha }(-\lambda _{\alpha }t^{\alpha })$
we have discovered a subdiffusive behavior for the all time regions rather
than only for long-time or short time regions that appears in the
literature. We have also obtained the exact solution for pdf's in the two
cases. For the first case, the exact solution of pdf is Gaussian. For the
second case, the exact solution of pdf is non-Gaussian, but it can reduce to
Gaussian pdf when $\alpha =1$.

\begin{acknowledgments}
K.S.F. acknowledges partial financial support from the Conselho Nacional de
Desenvolvimento Cient\'{\i}fico e Tecnol\'{o}gico (CNPq), Brazilian agency.
K.G.W. is pleased to acknowledge partial financial support received from the
Materials World Network Program and Metallic Materials and Nanostructures
Program of National Science Foundation, Washington, DC, under Grants No.
DMR-0710484.
\end{acknowledgments}

\appendix{}

\section{Integro-differential equation}

In order to obtain the integro-differential equation (\ref{eq10}) from Eq. (%
\ref{eq4a}) we employ the Fourier and Laplace transforms of the following
expressions:%
\begin{equation}
\mathcal{F}\left[ \frac{\partial ^{2}\rho (x,t)}{\partial x^{2}}\right]
=-k^{2}\rho (k,t)\text{ ,}  \label{eqA1}
\end{equation}%
and the convolution theorem for Laplace transform%
\begin{equation}
\mathcal{L}\left[ \int_{0}^{t}g\left( t-t_{1}\right) \rho (x,t_{1})\text{d}%
t_{1}\right] =g(s)\rho (x,s)\text{.}  \label{eqA3}
\end{equation}%
Firstly, we use the inverse Fourier transform and Eq. (\ref{eqA1}) in Eq. (%
\ref{eq4a}), and we find%
\begin{equation}
s\rho (x,s)-sg(s)\rho (x,s)-Dsg(s)\frac{\partial ^{2}\rho (x,s)}{\partial
x^{2}}=\left[ 1-g(s)\right] \rho (x,0)\text{ .}  \label{eqA4}
\end{equation}%
%
%
%
%
%
%
%
%
%
%
%
%
%
%
%
%
%
%
%
After both sides of Eq. (\ref{eqA4}) are divided by $s$, we reorganize it as
follows:%
\begin{equation}
\rho (x,s)-\frac{1}{s}\rho (x,0)-g(s)\rho (x,s)+\frac{1}{s}g(s)\rho
(x,0)=Dg(s)\frac{\partial ^{2}\rho (x,s)}{\partial x^{2}}\text{ .}
\label{eqA4-1}
\end{equation}%
Using Eq.(\ref{eqA3}) and applying the inverse Laplace transform in Eq.(\ref%
{eqA4-1}) we have 
\begin{equation}
\rho (x,t)-\rho (x,0)-\int_{0}^{t}g\left( t-t_{1}\right) \rho (x,t_{1})\text{%
d}t_{1}+\rho (x,0)\int_{0}^{t}g\left( t_{1}\right) \text{d}%
t_{1}=D\int_{0}^{t}g\left( t-t_{1}\right) \frac{\partial ^{2}\rho (x,t_{1})}{%
\partial x^{2}}\text{d}t_{1}\text{ .}  \label{eqA5}
\end{equation}%
%
%
%
%
%
%
%
%
%
%
%
%
%
%
%
%
%
%
%
Using the Leibnitz theorem for integral, we have the following relationship: 
\begin{equation}
\frac{\partial }{\partial t}\left[ \int_{0}^{t}g\left( t-t_{1}\right) \rho
(x,t_{1})\text{d}t_{1}\right] =g(0)\rho (x,t)+\int_{0}^{t}\frac{\partial }{%
\partial t}g\left( t-t_{1}\right) \rho (x,t_{1})\text{d}t_{1}\text{ .}
\label{eqA5-1}
\end{equation}%
The integral on the right side of Eq.(\ref{eqA5-1}) can be expressed as%
\begin{equation}
\int_{0}^{t}\frac{\partial }{\partial t}g\left( t-t_{1}\right) \rho (x,t_{1})%
\text{d}t_{1}=-\int_{0}^{t}\frac{\partial }{\partial t_{1}}g\left(
t-t_{1}\right) \rho (x,t_{1})\text{d}t_{1}\text{. }  \label{eqA5-2}
\end{equation}%
The integral on the right side of Eq.(\ref{eqA5-2}) can be integrated by
parts, and Eq.(\ref{eqA5-2}) can be further written as 
\begin{equation}
\int_{0}^{t}\frac{\partial }{\partial t}g\left( t-t_{1}\right) \rho (x,t_{1})%
\text{d}t_{1}=-g(0)\rho (x,t)+g(t)\rho (x,0)+\int_{0}^{t}g\left(
t-t_{1}\right) \frac{\partial }{\partial t_{1}}\rho (x,t_{1})\text{d}t_{1}%
\text{.}  \label{eqA5-3}
\end{equation}%
Substituting Eq.(\ref{eqA5-3}) into Eq.(\ref{eqA5-1}), Eq.(\ref{eqA5-1}) can
be rewritten as 
\begin{equation}
\frac{\partial }{\partial t}\left[ \int_{0}^{t}g\left( t-t_{1}\right) \rho
(x,t_{1})\text{d}t_{1}\right] =g(t)\rho (x,0)+\int_{0}^{t}g\left(
t-t_{1}\right) \frac{\partial }{\partial t_{1}}\rho (x,t_{1})\text{d}t_{1}%
\text{.}  \label{eqA5-4}
\end{equation}%
%
%
%
%
%
%
%
%
%
%
%
%
%
%
%
%
%
%
%

Firstly employing the operator $\partial /\partial t$ to Eq.(\ref{eqA5}) and
then substituting Eq.(\ref{eqA5-4}) into the equation, finally we have 
\begin{equation}
\frac{\partial \rho (x,t)}{\partial t}-\int_{0}^{t}g\left( t-t_{1}\right) 
\frac{\partial \rho (x,t_{1})}{\partial t_{1}}\text{d}t_{1}=D\frac{\partial 
}{\partial t}\int_{0}^{t}g\left( t-t_{1}\right) \frac{\partial ^{2}\rho
(x,t_{1})}{\partial x^{2}}\text{d}t_{1}\text{ .}  \label{eqA6}
\end{equation}

\end{document}